\documentclass[twocolumn,prl,amsmath,amssymb,floatfix,superscriptaddress,showpacs]{revtex4} 
\usepackage{graphicx}
\usepackage{dcolumn}
\usepackage{bm}

\usepackage{times}
\begin{document} 

\setlength{\topmargin}{0in}

\title{The Persistence and Memory of Polar Nano-Regions in a Ferroelectric Relaxor Under an Electric Field} 
\author{Guangyong Xu}
\affiliation{Physics Department, Brookhaven National Laboratory, Upton, New York 11973}
\author{P. M. Gehring}
\affiliation{NIST Center for Neutron Research, National Institute of Standards and Technology, Gaithersburg, Maryland, 20899}
\author{G. Shirane}
\affiliation{Physics Department, Brookhaven National Laboratory, Upton, New York 11973}
\date{\today}

\begin{abstract} 
The response of polar nanoregions (PNR) in the relaxor compound 
Pb[(Zn$_{1/3}$Nb$_{2/3}$)$_{0.92}$Ti$_{0.08}$]O$_3$ subject to a [111]-oriented 
electric field has been studied by neutron diffuse scattering.  Contrary to 
classical expectations, the diffuse scattering associated with the PNR 
persists, and is even partially enhanced by field cooling.  The effect of the 
external electric field is retained by the PNR after the field is removed.  
The ``memory'' of the applied field reappears even after heating the system 
above $T_C$, and cooling in zero field.
\end{abstract} 


\maketitle

Relaxor ferroelectric materials have garnered enormous attention in the 
materials science and condensed matter physics communities due to their 
record-setting dielectric and piezoelectric properties~\cite{PZT1}.  
They have supplanted lead zirconate (PZT) ceramics as the basis of 
state-of-the-art piezoelectric sensors and actuators that convert between 
mechanical and electrical forms of energy~\cite{Uchino}, and they show 
potential as ferroelectric nonvolatile memories~\cite{FEM_nature}.  
Compared to classic ferroelectrics, a unique property of relaxors is the 
appearance of  nanometer-sized regions having a local, randomly-oriented 
polarization at the Burns temperature $T_d$~\cite{Burns}, which is a few 
hundred degrees above the ferroelectric transition temperature (Curie 
temperature) $T_C$.  These polar nanoregions (PNR) act like precursors of 
the spontaneous polarization below $T_C$, and are believed to play a key role 
in the unusual relaxor behavior~\cite{Cross}.  Several theoretical models 
based on random fields in a dipole glass~\cite{Random_Field2,Random_Field1} 
have been proposed in order to explain the properties of the PNR. The PNR 
have been imaged directly with high resolution piezo-response force 
microscopy~\cite{Piezo_micro2,Piezo_micro1}, while extensive studies have 
been performed using both 
neutron~\cite{PMN_neutron3,PMN_diffuse,Xu_diffuse,Hiro_diffuse,PZN_diffuse3,
PMN_diffuse3} and x-ray~\cite{PMN_xraydiffuse,PMN_xraydiffuse2} diffuse 
scattering techniques.  These scattering measurements probe intensities in 
reciprocal space, which are Fourier transforms of real space structures, and 
provide valuable information about the magnitudes and orientations of the 
polarizations of the PNR, their dynamic properties, and the sizes and shapes 
of the PNR. 

There are many unsolved fundamental questions concerning the PNR.  Do they 
facilitate the ferroelectric phase transition of the system, or do they impede 
it? What prevents them from dissolving into the surrounding, macroscopically 
polar, environment below the ferroelectric phase transition temperature $T_C$? 
In studies of ferroelectrics, an external field is often applied to monitor how 
the system responds.  Yet few studies have examined the PNR response to an 
external field.  Previous studies~\cite{PMN_efield,PZN-8PT} have shown that the 
neutron diffuse scattering measured in directions transverse to the scattering 
vector can be partially suppressed in relaxor systems by an electric field.  
In particular the diffuse scattering measured transverse to the (300) Bragg 
peak in PZN-8\%PT is unaffected by an electric field applied along the [001] 
direction, whereas that transverse to the orthogonal (003) Bragg peak is 
significantly suppressed.  The anisotropic response of the diffuse scattering 
in this case seems to suggest that those PNR having "properly oriented" 
polarizations could be forced to "melt" into the surrounding polar environment 
with the assistance of an external field, thereby producing a more 
"microscopically-uniform" polar state.

\begin{figure}[ht]
\includegraphics[width=\linewidth]{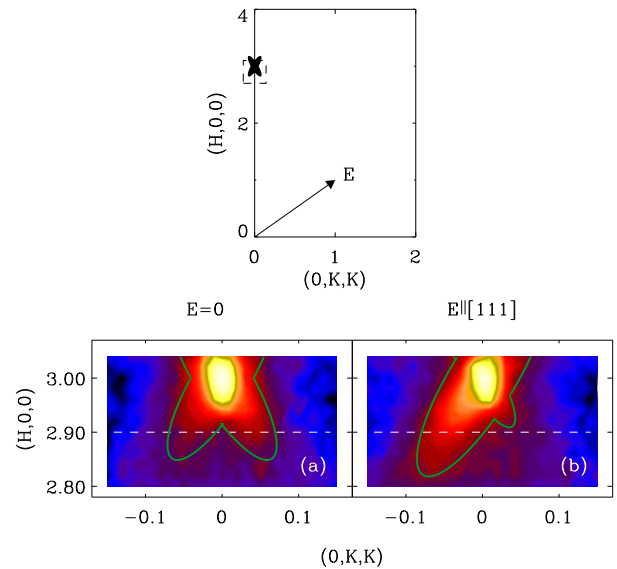} 
\caption{The top frame is a schematic of the (HKK) reciprocal scattering plane, 
in which neutron diffuse scattering measurements were performed close to the 
(300) Bragg peak.  The electric field was oriented along [111] as shown by the 
arrow.  The bottom frames show results from measurements made at $T=300$~K 
after the sample was zero-field cooled (ZFC) (a) and field-cooled (FC) (b) 
through 
$T_C$.  The dashed lines indicate the location of the cuts at (2.9,K,K), shown 
later in Fig.~\ref{fig:3}. The solid green lines are guides to the eye to help 
emphasize the symmetric (a) and asymmetric (b) ``butterfly'' shapes of the 
diffuse scattering.}
\label{fig:1}
\end{figure}

One of the most studied relaxor systems is the lead perovskite 
Pb(Zn$_{1/3}$Nb$_{2/3}$)O$_3$ (PZN). When mixed with PbTiO$_3$ (PT) to form 
solid solutions, PZN-$x$PT single crystals exhibit ultrahigh piezoelectric 
responses~\cite{PZT1,PZN_phase1}. The piezoelectric property is optimal around 
$x=8\%$, which is the PT concentration of the single crystal
examined in our study.  The zero-field structure~\cite{Noheda} and the 
[001]-oriented electric-field 
induced structural changes of PZN-8\%PT have been studied in 
detail~\cite{PZN_efield,PZN_field}. The lattice parameter in the cubic phase 
is $a=4.045$~\AA.  Our experimental measurements are then described in terms of 
reciprocal lattice units (r.l.u.) where 
1~r.l.u.~$= a^*=2\pi/a=1.553$~\AA$^{-1}$.
Because PZN-8\%PT has a rhombohedral ground state 
with $\langle111\rangle$ type polarizations at low temperatures, we chose to 
investigate the effect of an electric field applied along the [111] direction.  
Contrary to classical expectations, we find that the diffuse scattering does 
not diminish with the application of an external electric field. Instead, 
after field cooling, only a part of the diffuse scattering is suppressed in
reciprocal space, while another part is markedly enhanced.  We also observed 
interesting ``memory'' behavior in which the effects of an external field 
remain after the removal of the field, vanish above $T_C\sim450$~K, 
and then reappear after zero-field cooling (ZFC) below $T_C$.

The PZN-8\%PT single crystal used in this study was provided by TRS 
ceramics~\cite{Support_prl}.  
The crystal is  rectangular in shape, having dimensions 
$5\times5\times3$~mm$^3$ with (111), ($\bar{2}$11), and (0$\bar{1}$1) surfaces. 
The Cr/Au electrodes were sputtered onto the top and bottom (111) crystal 
surfaces.  
The neutron diffuse scattering measurements were performed with the BT9 
triple-axis spectrometer located at the NIST center for Neutron Research.  The 
measurements were made using a fixed incident 
neutron energy $E_i$ of 14.7~meV, obtained from the (002) reflection of a 
highly-oriented pyrolytic graphic (HOPG) monochromator, horizontal beam 
collimations of 40'-40'-40'-80', and the (002) reflection of an HOPG analyzer 
to fix the energy of the scattered neutron beam.  The sample was 
oriented with the (0$\bar{1}$1) 
surface facing vertically.  The scattering plane is therefore the (HKK) plane, 
which is defined by the two primary vectors [100] and [011].  The [111] 
electric field direction lies in the horizontal scattering plane.

\begin{figure}[ht]
\includegraphics[width=\linewidth]{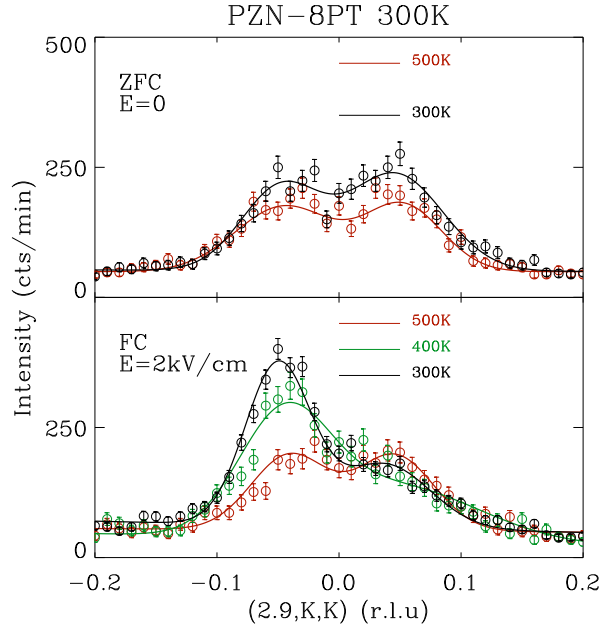} 
\caption{Linear scans performed at (2.9,K,K), for the ZFC state (top panel),  
and the FC state (bottom panel), measured at different temperatures.  The solid 
lines are fits to a sum of two Gaussian peaks.}
\label{fig:3}
\end{figure}

Ideally, diffuse scattering intensities should be measured around different 
\{111\} peaks to gauge the net effect of a [111]-oriented electric field. 
However, the neutron diffuse scattering is very weak around all of the \{111\} 
peaks due to a small neutron scattering structure factor~\cite{Hiro_diffuse}. 
Instead, the measurements were performed near the (300) 
Bragg peak, where the diffuse scattering is strong, and the Bragg peak 
intensity is weak, as shown schematically in the top frame of Fig.~\ref{fig:1}. 
Because of limitations on the maximum achievable scattering angle, we were only 
able to 
measure the lower half ($H \le 3$) of the butterfly-pattern.  The data obtained 
at $T=300$~K ($< T_C \approx 440$~K in zero field) are shown in the bottom 
frame of Fig.~\ref{fig:1}.  When the crystal is zero-field cooled (ZFC), the 
diffuse scattering pattern is symmetric about the [100] axis, and forms the 
butterfly-shape with ``wings'' of equal intensity on each side shown in 
Fig.~\ref{fig:1}~(a).  This pattern is consistent with the three-dimensional 
diffuse scattering distribution consisting of $\langle110\rangle$  rod-type 
intensities reported in Ref.~\onlinecite{Xu_3D}, where PNR with different 
polarizations contribute to different $\langle110\rangle$ diffuse rods.  
Because of the non-zero out-of-plane wavevector resolution of our measurements, 
the tails of the $\langle110\rangle$ rods are picked up when scanning away from 
the center of the Bragg peak in the (HKK) plane.  This results in the 
butterfly-shaped pattern, which (neglecting the ${\bf Q}^2$ factor in the 
diffuse intensity) is symmetric about the wavevector ${\bf Q}$=(3,0,0), as 
required by the crystal symmetry.

After zero-field cooling to $T=300$~K (below $T_C$) we applied a moderate 
electric field $E=2$~kV/cm along the [111] direction.  No change in the diffuse 
scattering was observed, suggesting that the PNR are robust against an 
external field in the ferroelectric phase.  The sample was then heated to 
$T=500$~K, and an electric field of $E=2$~kV/cm along [111] was reapplied.  
No changes were evident at this temperature as a result of the field.  
However, after field-cooling through $T_C$  ($\approx 460$~K for $E=2$~kV/cm) 
back to $T=300$~K, the diffuse scattering pattern developed a strong asymmetry 
about the [100] direction as shown in Fig.~\ref{fig:1}~(b).  The left wing is 
clearly enhanced while the right one is suppressed.  A more detailed look at 
the temperature dependence of the diffuse scattering is provided in 
Fig.~\ref{fig:3}.  Here characteristic linear scans of the diffuse scattering 
intensity measured along (2.9,K,K) (see the dashed line in Fig.~\ref{fig:1}) 
are plotted.  In the ZFC state the symmetric double-peaked profiles indicate 
that the intensities from both wings increase equally with cooling.  In the FC 
state, for which the [111]-oriented $E$-field was initially applied at 
$T=500$~K in the cubic (paraelectric) phase, it is only after cooling below 
$T_C$ that the intensity of the left wing begins to increase, whereas the right 
wing intensity remains almost constant.  At $T=300$~K, the left wing intensity 
is clearly much higher than that measured in the ZFC state.  Upon removal of 
the electric field at $T=300$~K, 
the asymmetric diffuse pattern still remains, indicating that the 
PNR are ``locked in'' the FC configuration.  This behavior is quite similar to 
that of a spin glass system~\cite{SpinGlass} where cooling through $T_C$ in 
zero and non-zero magnetic fields leads to different final states, for which 
the difference is only observable below $T_C$.

\begin{figure}[ht]
\includegraphics[width=\linewidth]{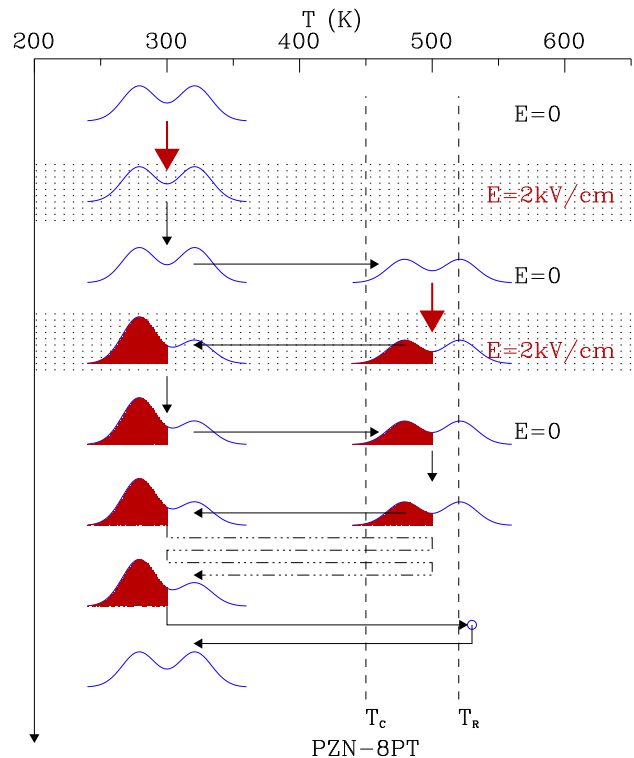}
\caption{A schematic diagram showing the chronological change of the diffuse 
scattering from the PNR, described by the two peaks at (2.9,K,K).  The shaded 
peaks denote states where the PNR retain a ``memory'' of the electric field. 
The dashed line at $T_R$ indicates the temperature above which this field 
memory is lost.  Red arrows are used to denote the onset of the external 
electric field.}
\label{fig:4}
\end{figure}

Additional measurements reveal an extremely interesting memory effect.  After 
field-cooling below $T_C$, and then removing the field (for which the leads 
across the sample were shorted), the sample was heated back up to the cubic 
paraelectric 
phase at $T=500$~K, and kept there for $\agt 2$~hours.  At $T>T_C$, the diffuse 
scattering intensities measured at (2.9,K,K)  consist of two symmetric peaks.  
However, the asymmetric lineshape reappears after cooling back to $T=300$~K.  
The effect of the previous electric field is still "remembered" after extensive 
thermal cycling through $T_C$.  A schematic representation of the diffuse 
intensity lineshapes measured at (2.9,K,K) is shown in Fig.~\ref{fig:4} in 
chronological order.  This memory effect is lost when the compound is heated 
up to $T=525$~K and cooled in zero field.  The highest temperature at which 
the compound can still retain the memory of previous [111] electric field thus 
lies between 500~K and 525~K, roughly 50~K greater than $T_C$, and is denoted 
as $T_R$, the ``repoling'' temperature, in Fig.~\ref{fig:4}.  Similar 
``repoling'' effects have also been observed in other relaxor systems such as 
PLZT~\cite{PLZT_memory} and SBN~\cite{SBN_memory}. 

The persistence of the diffuse scattering in this relaxor compound in the 
presence of an external field is truly remarkable.  In a normal "classic" 
ferroelectric compund an external electric field promotes the establishment of 
a macroscopic ferroelectric phase via domain rotation.  It was similarly 
expected that an external electric field would be able to affect the PNR such 
that they would tend to align with the field and merge into the surrounding 
ferroelectric polar lattice, resulting in a decrease of the diffuse scattering 
intensity.  Our results clearly indicate otherwise.  The enhancement of diffuse 
scattering demonstrates that a more uniform polar state has not been achieved.  
The low temperature phase of PZN-8\%PT consists of PNR embedded within a 
ferroelectric polar environment.  This is a very interesting situation in which 
a stable short-range polar order (the PNR) persists through a phase
transition and coexists with long-range polar order.  
The application of an external electric field does not suppress the short-range 
polar order.  Instead, the short-range polar order develops inside the 
long-range polar order, and can even be "enhanced" by the field.

There have been theoretical arguments~\cite{Random_Field1} based on 
random-field models suggesting that the low-temperature glassy phase of the 
relaxor may not be able to switch to the normal behavior of ferroelectrics 
through poling.  Our results provide unambiguous confirmation of this thesis 
in that, even after cooling under a moderate external electric field, the 
relaxor PZN-8\%PT still exhibits the strong neutron diffuse scattering 
characteristic of embedded PNR, and thus does not transform into a uniform 
ferroelectric phase below $T_C$.  Contrary to the naive belief that 
ferroelectric domains develop from the PNR, these unusual local polar 
nanoregions appear to be displaced, or "out-of-phase" with respect to the 
surrounding lattice.  In addition to the already well-known 
``uniform-phase-shift'' concept~\cite{PMN_diffuse}, the PNR can also have 
polarizations~\cite{Xu_3D} that are different from that of the surrounding 
polar environment.  Neither a moderate external field, nor the ferroelectric 
polar environment itself, has the energy required to dissolve these 
out-of-phase polar nanoregions.  This novel polar nanostructure clearly 
distinguishes relaxors from classic ferroelectrics.

An interesting speculation that can be made from our results is that the 
diffuse scattering intensity is redistributed, or shifted from one wing of the 
butterfly to the other.  This may suggest that the polarizations and shapes of 
the PNR are not different in the FC and ZFC states.  Instead, there may be a 
redistribution of PNR with different polarizations.  If we make the assumption 
that the PNR retain the $\langle110\rangle$-type polarizations after 
field-cooling,  the asymetric diffuse scattering pattern observed in the 
FC state can be explained by a simple model.  Knowing that the diffuse 
scattering consists of $\langle110\rangle$ rods~\cite{Xu_3D}, we can see that 
the lower left wing of the butterfly shown in Fig.~\ref{fig:1}~(a) originates 
from the tails of rods of diffuse scattering intensity oriented along the 
[110] and [101] directions, and the lower right wing originates from the tails 
of rods oriented along the [1$\bar{1}$0] and [10$\bar{1}$] directions.  More 
importantly, the polarizations of PNR contributing to different 
$\langle110\rangle$ diffuse rods have also been clearly identified to point 
along the perpendicular $\langle1\bar{1}0\rangle$ directions~\cite{Xu_3D}.  
Therefore, the enhanced left wing observed in our measurements arises from 
PNR with polarizations along [1$\bar{1}$0] and [10$\bar{1}$], both 
perpendicular to the external [111] electric field,  and the suppressed right
wing arises from PNR with polarizations along [110] and [101], which are not 
perpendicular to the field.  There are then two possible scenarios that can 
lead to the enhancement of PNR with polarizations perpendicular to the field. 
The first is one in which the PNR are affected directly by the field so that 
those PNR having perpendicular polarizations are more favored by the FC 
process.  This is quite unlikely since the electric field tends to align 
electric dipole moments along the field direction.  The second is one in which 
the PNR are not affected directly by the external field.  In the paraelectric  
phase, they can have polarizations that point along any one of the six 
$\langle110\rangle$ directions and the field has no effect.  In the low 
temperature ferroelectric phase, however, the six $\langle110\rangle$ 
directions are no longer equivalent.  The PNR may tend to grow with 
polarizations perpendicular to the  $\langle111\rangle$ type polarization of 
the surrounding lattice.  This actually prevents them from merging into the 
surrounding polar lattice.  The electric field only affects the alignment of 
the ferroelectric domains in which the PNR are embedded, so that the 
enhancement of perpendicular PNR can be observed at a macroscopic level. This 
model is purely speculative, and more experimental support is required to 
confirm or refute it.  However its simplicity can be appreciated as a starting 
point for developing an understanding of PNR behavior.

After removal of the field, the effect on the crystal remains, and so will the 
configuration of the PNR.  However we do not currently 
understand the reason for the ``memory'' effect, where heating to 
$T_C < T < T_R$ does not ``depole'' the system. The memory of the electric 
field is only hidden, and reappears again when zero-field cooled  below $T_C$.  
Considering the case of magnetic systems with random fields, the effect of an 
external magnetic field can lead to very different behaviors, even a change 
of $T_C$, between FC and  ZFC~\cite{Birgeneau}.  There nonequilibrium effects 
play an essential role.  In relaxor systems, similar random-field models for a 
dipole glass can be considered.   Based on our measurements of the structural 
changes~\cite{Support_prl}, $T_C\sim 460$~K for $E=2$~kV/cm was only slightly 
higher than $T_C\sim440$~K for $E=0$~kV/cm.  However, the electric field may 
also have effects on some other hidden degrees of freedom that do not fully 
dissolve at $T_C$, and therefore are not directly observable in structural 
measurements.  For example, one such possibility is that the effect of the 
electric field may be ``remembered'' by the lattice dynamics, i.\ e., atomic 
motions associated with the PNR.  Even when the whole system goes into a 
paraelectric phase above $T_C$, the memory of the previous configuration could 
be embedded within the atomic motions in a metastable phase at $T_C<T<T_R$.  
The system may not be able to equilibrate over experimentally accessible time 
scales. These hidden effects could resurface again upon cooling and produce the 
spontaneous ``repoling''.  The results we have reported here also have 
significant application implications in areas such as ferroelectric memory 
devices.  An electric field can be used to embed information in the relaxor 
system.  After removing the field, the information will only surface for 
$T<T_C$, but can not be erased unless heating the system up to $T>T_R$, making 
the device useful for temperatures  in the neighborhood of $T_C$, where the 
various relaxor properties are optimal.

We would like to thank C. Broholm, H.~Hiraka, S.~M.~Shapiro, C.~Stock, 
J.~M.~Tranquada, and Z.~Zhong,
for stimulating discussions. Financial support from the U.S. Department of 
Energy under contract No.~DE-AC02-98CH10886 is also gratefully acknowledged.


\begin{thebibliography}{29}
\bibitem[{Sup()}]{Support_prl}
\bibinfo{note}{For this particular sample, the structure and therefore the
  phase diagram was determined by monitoring the intensity and splitting of
  (200), (022) and (111) Bragg peaks using elastic neutron scattering
  techniques. All measurements were performed with cooling processes. In zero
  external field $T_C \approx 440~$K for the transition from cubic to
  tetragonal symmetry, and $T_{C2} \approx 340$~K for the transition from
  tetragonal to rhombohedral symmetry. Under an external electric field
  $E=2$~kV/cm oriented along the [111] direction, $T_C \approx 460$~K, slightly
  higher than the zero-field value. $T_{C2} \approx 340$~K does not change much
  under field.}
\end{thebibliography}

\end{document}